\documentclass[aps]{article}
\usepackage{geometry}                
\usepackage{graphicx}

\usepackage{amssymb}
\usepackage{epstopdf}
\usepackage{amsmath}

\usepackage{amsfonts}
\usepackage{bm}
\usepackage{hyperref}

\DeclareMathAlphabet{\mathpzc}{OT1}{pzc}{m}{it}

\title{Thermalization of Gases: A First Principles Approach}
\author{Clifford Chafin\\\ \small{Roto9 Energy, Chapel Hill, NC 27695}\thanks{cechafin@ncsu.edu}}

\begin{document}
\maketitle
\begin{abstract}
Previous approaches of emergent thermalization for condensed matter based on typical wavefunctions are extended to generate an intrinsically quantum theory of gases.  Gases are fundamentally quantum objects at all temperatures, by virtue of rapid delocalization of  their constituents.  When there is a sufficiently broad spread in the energy of eigenstates, a well-defined temperature is shown to arise by photon production when the samples are optically thick.  This produces a highly accurate approximation to the Planck distribution so that thermalization arises from the initial data as a consequence of purely quantum and unitary dynamics.  These results are used as a foil for some common hydrodynamic theory of ultracold gases.  It is suggested here that strong history dependence typically remains in these gases and so limits the validity of thermodynamics in their description.  These problems are even more profound in the extension of hydrodynamics to such gases when they are optically thin, even when their internal energy is not low.  We investigate rotation of elliptically trapped gases and consistency problems with deriving a local hydrodynamic approach.  The presence of vorticity that is ``hidden'' from order parameter approaches is discussed along with some buoyancy intrinsically associated with vorticity that gives essential quantum corrections to gases in the regimes where standard perturbation approaches to the Boltzmann equations are known to fail to converge.  These results suggest that studying of trapped gases in the far from ultracold regions may yield interesting results not described by classical hydrodynamics.  
\end{abstract}

\section{Introduction}
Thermodynamics has been of fundamental importance to physics, chemistry and engineering since at least the 19th century.  Nevertheless the meaning of temperature and entropy was elusive until the classical theory of gases and, with the arrival of quantum mechanics, one could argue that they are once again mysterious.  We certainly have formal partition functions to generate predictions for quantum statistical mechanics yet the reasons for their successes have remained obscured.  They arose as natural analogs to the classical calculations but, unlike the classical case, lacked a reason for the system to gravitate to the implied distribution of eigenstates or be sensibly described by an ensemble at all.  

The boundary between classical and quantum behavior has never been well-defined.  It is not even clear if there is a meaningful transition.  Initially it was thought that quantum behavior was limited to very small bodies.  The appearance of superfluidity and superconductivity made it clear that this was not the case.  The ``classical limit'' has never been clearly understood.  Even the suggestion of it implies that quantum behavior will naturally lead to a classical looking world for large enough collections of atoms at high enough temperature (or internal energy).  None of this is clear and attempts to reconcile this include decoherence \cite{Schlosshauer}, small nonunitary evolution, dynamic typicality and the eigenstate thermalization hypothesis \cite{Popescu,Srednicki,Srednicki1, Bartsch}.  Important progress has been made in nonequilibrium classical statistical mechanics in recent years \cite{Jarz,Cohen} yet a quantum analog has been elusive.  

This author has pioneered a picture of classical matter that has a well-defined quantum description with fully Schr\"{o}dinger evolution at all times and scales.  It is, however, very special in that it has well-defined macroscopic shape, orientation, and atomic locations for condensed matter \cite{Chafin-pip-meas,Chafin-I} unlike the ground states of any interesting Hamiltonian.  This is powerful in that it leads naturally to quantum measurement statistics and a many worlds-like partitioning of the space for long times.  The origin of such a special state is somewhat unclear but some qualitative arguments are provided on how this could arise for condensing gases in a photon poor space.  Dual to this is a picture of QED \cite{Chafin-pip-qed} that is built on a tower of many photon spaces in real space coordinates with a many time picture that agrees with QED on the equal times diagonal.  Some complications arise in the shape of the domain of dependence in this many times space but the equal times diagonal is clearly contained in it.  

One important phase that does not resolve easily from this picture is gases, specifically, we desire a classical limit for the theory of gases.  The condensed matter objects have long lasting persistent macroscopic features because the delocalization times are very long.  For gases, the particles delocalize rapidly so that no such picture is reasonable.  This is especially important due to the recent work on ultracold gases and the attempt to attach hydrodynamic and thermodynamic quantities such as viscosity and temperature to them \cite{Dolfovo, Son}.  Without a truly quantum understanding of classical gases, it is unclear when these notions are well-defined.  

Gases occupy an especially important place in our understanding and history of heat and hydrodynamics.  The ballistic and statistical collision model put forward by Maxwell and Boltzmann led both to a microscopic understanding of heat and entropy and to a perturbative picture of hydrodynamics from them.  Some problems still remain such as how to handle higher order perturbations in a convergent and sensible manner \cite{Dorfman, Cohen}.  In the quantum case, it is not clear how such a ``3D'' picture can ever arise from such an intrinsically delocalized object.  Discussions often leap over this with some vague quantum assertions involving the relative scale of deBroglie wavelengths and interparticle separation \cite{Tolman}.  The approach here will be to consider the usually neglected role of the photon in thermalization of matter.  

It is known that photons will not equilibrate to a Planck distribution without matter but it seems to be unsuspected that photons are essential in the equilibration of quantum matter to lead to the kinds of eigenstate distributions that dominate the microcanonical ensemble.  This article is to demonstrate that this is actually the case and so resolve both thermalization of gases and give the first truly dynamical explanation of the Planck distribution.  It has been an enduring mystery as to how microscopic dynamics of gases can lead to this and, neglecting the role of the many photon number states and the nontrivial subset of realistic initial data, one is naturally led to the conclusion that a gaseous star should not give such a distribution.\footnote{This has motivated the notion that there is a continuum of radiating states in the sun's surface akin to a metal  \cite{Robitaille} despite that fact that the surface plasma contains a large fraction of undissociated gas.}  A surprising conclusion of this article will be that we have been overly hopeful in our assumptions that classical gas kinetics provides a real derivation of the behavior of high temperature gases.  We will see these are intrinsically quantum objects at all temperatures and require a derivation that respects this without any reference to the billiard kinetics of the 19th century.  

\section{``Classical'' Gases and the Planck Distribution}

Why is thermalization of quantum systems more problematic than the classical gas case?  Classically we describe the system as a single point moving on a submanifold in $\mathbb{R}^{6}$ phase space defined by the conserved quantities of energy, angular momentum, etc.  Ergodicity does not resolve the macroscopic averaging problem due to long times to traverse this manifold but most states drift rapidly to a condition with well defined parcel averaged local density, velocity, temperature, etc.\ \cite{Villani} that agrees with the submanifold averages for the vast majority of times.  Even defining evolving systems in the classical case is a ``battle of scales'' whereby we must partition the system into parcels large enough to have macroscopic meaning for thermodynamic variables but small enough to allow smooth variation of these quantities between them.  To handle the case of strong gradients when these conditions are valid we have only perturbative approaches and they have met with poor success \cite{Dorfman}.  Determinism in such a system in macroscopic variables is certainly false as is evident by the statistical nature of turbulence and importance of thermal fluctuations.    Nevertheless, in their domain of applicability, classical kinetics works well and provides an intuitively appealing approach for students.  

In the quantum gas state, eigenstates never thermalize.  Distributions made of eigenstates maintain their exact relative phase contributions.  The microcanonical ensemble approach suggests that only states with nearly constant energy eigenstates should contribute.  This is especially problematic for cases with hydrodynamic and thermal gradients since there is no evident way to ``partition'' a many body wavefunction into local parcels as in the classical case.  

Our central thesis is that thermalization in gases in the ``classical limit'' depend on 1. an internal photon field that drives equilibration over a tower of different photon number states and 2. an initial photon free distribution of eigenstates that has much larger energy spread, $\Delta E$, than mean energy $<E>$.  The many body currents driven by the spread in eigenstates can then drive amplitude between various photon number states until they become relatively small.  Not surprisingly, this means the optical depth of the sample is very important in equilibration.  For gases that are not optically thick we can expect radiation that is typical of the discrete atomic levels of the atoms and some variation in the Maxwell-Boltzmann distribution that is related to the relative energy density of the atoms versus the photon field.  This is certainly expected to be small but, for photon driven processes, possibly including nucleation and evaporation, a change in this field may be important.  

It is important to find conditions that lead to some sort of universality at the macroscopic level.  For starters, consider the case of no thermal gradients or hydrodynamic flow.  If we chose an eigenstate then the velocity distribution of the individual particles can be whatever we choose but the overwhelming number of states give M-B when the energy is such that the wavelengths are smaller than the interparticle separation.  Interestingly, this leads to no density oscillations, hence no photon production.  This does not mean the electron orbitals are undistorted.  At the n-body diagonals in the atomic center-of-mass coordinates there are density variations that are reflected in the electronic orbitals however the state is still a net eigenstate so no photons are produced.\footnote{This is in contrast with the often vague assertion regarding QED ``fluctuations'' driving excited eigenstates to decay via photon emission as in the case of the 2p$\rightarrow$1s transition despite that such QED corrections should simply shift the eigenstates.  This author's opinion is that all such transitions require some external perturbation and that it is practically impossible to ever create a pure 2p state anyway.}

This perspective on thermalization is built on the notion of a ``typical wavefunction'' \cite{Chafin-II} so that nature is always in a ``pure state'' in the language of ensemble based statistical mechanics.  Consider an eigenstate of a noninteracting gas of net energy $E$.  Locally we can describe the gradient of oscillations as having a well-defined wavevector $\tilde K$ in the 3N dimensional configuration space (suppressing spin labels).  For a random such many body vector we can locally project on the $N$ 3D subspaces to get a set of 3-vectors $k_{1},k_{2}\ldots k_{N}$.  By a M-B distribution we mean that this set of vectors almost always correspond to the momenta of the M-B distribution.  Note that we have not required that the net state be a plane wave or that the system have any localization properties and the corresponding ``uncorrelated'' property of a classical billiard ball gas.  Now let us consider more realistic wavefunctions and their evolution. 

If we choose a superposition of two material eigenstates with energy difference $\Delta E$ then there are density fluctuations with frequency $\omega=\Delta E/\hbar$ that can drive photon production and drive the amplitude from the no-photon space to the many photon spaces where each increasing level has less energy apportioned to the atomic motions and more to the photons.  This still cannot lead to some universal thermalization since once we reach the state\footnote{The notation used here is the same as in Chafin 2015 \cite{Chafin-pip-qed}.} $\Psi_{N,m}$ of $N$ atoms and $m$ photons with $m \hbar \omega=\Delta E$ there is no higher photon state that can be reached and we certainly do not have a Planck photon distribution!  To see in more detail what occurs during this process note that the oscillations are all initially of frequency $\omega$ so that, regardless of that the atomic energy spectra is, the period of the charge currents are also $\omega$.  If this is far from one of the atomic frequency lines, the efficiency of this process will be small because the atoms will be effectively ``stiffer'' against such oscillations.  The resulting norm flux is $\Psi_{N,0}\rightarrow\Psi_{N,1}$.  The photon field now can interact with the atoms and be absorbed giving a corresponding downwards flux of norm.  The new state still has material oscillations at the frequency $\omega$ and continues to drive norm up to higher photon number states.  Note that the usual restriction of the atomic radiative frequencies have not played any role here.  Line broadening is know to exist from Doppler, N-body and proximity effects but here we are simply driving the full set of charges at a given frequency and therefore photons are produced at that frequency regardless of the individual atoms sprectra.  

Let us now consider a state with no initial photons and broad spread in eigenstate energy.  The current fluctuations will drive photon production up the tower of states $\Psi_{N,0}, \Psi_{N,1}, \Psi_{N,2}\ldots$ until some equilibrium is reached.  The flux of norm out of or in to any one state is minimized when the energy distribution of the atomic and photon parts of the wavefunctions vanish.  For a system in a confined box of length $L$ this gives a maximum state $\Psi_{N,p}$ where $p=E/\hbar\omega_{0}$ and $\hbar \omega_{0}=3 h^{2}/8mL^{2}$.  There is no reason to believe that the evolution will completely fill all of these state with equal norm for any such initial distribution and that there will not be some dynamical variation in these fillings over time.  However, it is our hypothesis that such an assumption is very accurate for most times after some equilibration time defined by the rate that the fluctuations in each $\Psi_{N,s}$ state reach the two-body center-of-mass diagonals where the radiation can then be generated by the relative electron-nuclei coordinates.  Such a situation gives nearly equal norm to all states $\psi_{N,s}$ and a nearly uniform weighted energy distribution at each many body point $\tilde X$, $E(\tilde X)=\hat E(\Psi)/\mathcal{N}(\Psi)$, for each one.  The state $\Psi_{N,0}$ will have no photon energy and the states with $\Psi_{N,s\rightarrow\infty}$ will have almost all energy in the photons.  

If the photons and matter are only weakly coupled we can view the matter and photon eigenstates as nearly spanned by the products of eigenstates of each.  Such a ``quiescent limit'' is well described by a tower of material and photon eigenstates chosen so the net norm of each (as defined by \cite{Chafin-pip-qed}) is uniform.  Since the thermal energy density of photons is almost always extremely small compared to the material contributions, a condition we call ``matter dominance,'' this lets us consider the physically narrow window of states where the material part of the wavefunctions have nearly the same energy.  
If we were to have a distribution of energies in a single photon number state, there would be oscillations in the fields that drive dipolar forces on the atoms so we expect spreading in photon number until the material and photon parts of each $\Psi_{N,p}$ are nearly eigenstates.  
This gives the usual microcanonical weighted distribution of states here.  The photons' components in this range of states are dominated by nearly equal energy states of different photon number.  The dominant contributions to each photon number space are then the same as in the Planck distribution so we generally get an excellent approximation to it.  This is done in a fashion that works for optically thick gases and requires no ``cavity'' or ``absorbing bodies'' to drive the equilibration.  This condition leads to a set of states that are described by microcanonical ensembles in that only the equal energy many body eigenstates need to be considered.  By this, we mean that each $\Psi_{N,p}$ is a near eigenstate that is likely to have macroscopic averaged values consistent with the vast majority of such states so that taking the ensemble average gives the same result.  

One of the interesting features of this model is that it allows for derivations of transition times to equilibrium and deviations from the Planck distribution.  When the ``matter dominance'' condition is not strongly obeyed, as in a very dilute but still optically thick and cold gas, there should be variations in the photon distribution due to correlations with the material states.  It should be noted that this distribution is transient just as thermal distributions are for most classical cases.  Unlike the eigenstate case with a M-B distribution, there are real temporal fluctuations present here although it is not clear if they are large enough to measure or are in any sense universal.  The initial energy spread of the initial data and the size of the cavity of the experiment and possibly even the duration of the equilibration over the many photon spaces may play a role in this.  We should carefully distinguish such temporal from spatial fluctuations.  Spatial ones can exist in a persistent fashion due to the effects of two-body potentials and the induced correlations.  One concern immediately is, if temporal fluctuations are not universal, how can Brownian motion be observed and exist as a universal phenomenon?  Just as in the case of quantum measurement, this involves the interaction of a classical body, with well-defined and long lasting localization, with an intrinsically quantum object, a gas.  For the same reasons as in the theory of quantum measurement associated with this line of thought \cite{Chafin-pip-meas}, an observer will only see the body at a well-defined point even though it is subjected to delocalizing actions correlated with the gas interactions.  Each of these questions gives important directions for future work.  

\section{Ultracold Gases}

One of the motivations for this investigation was the recent experimental work in ultracold atomic and molecular gases \cite{Dolfovo}.  It has been observed that there are oscillatory ``scissors'' and ``breathing'' modes of such gases in harmonic traps that seem well described by Euler's hydrodynamics.  Damping of such modes is also observed and it is suggested that hydrodynamics are valid for these systems and that some universal quantum bound on viscosity \cite{Son} is approached by these systems.  There has been significant work attempting to establish a phase portrait of fermions at this very low temperature on the assumption that there is some universality that will scale up to such diverse systems electrons in metals and nucleons in neutron stars \cite{Dolfovo}.  Without a quantum understanding of classical gases that exists at the level of a typical wavefunction, we are saddled with the usual formal ensemble based approaches with no justification of the limits of their validity.  Furthermore, we are left wondering if the ``large N'' limit is a sufficient justification for assigning temperature and hydrodynamic variables to such clouds which we have reason to believe are described by general many body wavefunctions at all times.  

The reasons for concern here is that these clouds are ``cooled'' evaporatively but we do not expect that the wavelength be small compared to the interparticle separation so we have even less reason to be satisfied with the billiard ball inspired arguments for such systems.  Additionally, these clouds are optically thin and the time for photons to play a role in equilibration are very long so thermal equilibration may not be a property such clouds possess.  Furthermore, we expect the background radiation to be dominated by that of the much warmer walls of the chamber.  It seems likely that the matter currents in these clouds never subside and we are left with superpositions of different energy eigenstates that preclude them being described by a well-defined temperature.  Of course, the same cooling by evaporation by lowered trap barriers will reduce both the standing wave energy and the fluctuations simultaneously.  Fluctuations in currents above the threshold will also evaporate so some reduction in these fluctuations, certainly radial to the trap edge, is expected.  In terms of the distribution of eigenstates, for a spherically symmetric cloud with no net angular momentum, such a state will tend to $\Delta E\approx0$ so that this is one case where we tend to an eigenstate so that thermalization is reasonable to expect \cite{Houcke}.  This is a special case where temperature based on the microcanonical and global approach to the system makes sense.  They do not ``autothermalize'' without radiation but have been coaxed to such a state through evaporation and varying the interaction strength.  This temperature is a globally defined quantity and it is not clear that localizing this quantity will have meaning for more general clouds.  Near the zero crossing of the scattering length, equilibration does not happen so the faith of so many in thermo and hydro treatments of such clouds is perplexing.  It is certainly suggestive that the history of the preparation of the system is likely to persist to some extent in these clouds, in contrast with our implicit expectations of thermodynamic systems.  

Trap losses are driven by three body recombination which reminds us that these clouds are really excited eigenstates and the true ground states are solid clusters of the typically metal atoms that they contain.  These losses can be kept to low levels but this three body effect is then also forbidden as an effective means of thermal equilibration.  Photons from external sources of sufficiently low energy provide a means of equilibrations but also introduce heating effects.  In general we have few ways to probe such gases and measurements of them are often destructive optical projection events.  Recent progress \cite{Zwierlein} has enable nondestructive measurements but enforcing changes on the clouds is still limited to a few crude tools including trap changes and optical imprinting of bulk phases.  

Not only are the thermal wavelengths of such ultracold gases typically long compared to the interparticle separation but the interactions are often tuned to ``unitarity'' whereby the two body scattering length diverges.  For many body systems this means that the curvature associated with the near (quasi)ground state of the cloud gives large regions of repulsion and curvature that is not associated with heating.  This is often interpreted as forcing ``correlated'' motion in that the two body correlation functions are strongly constrained even when spins are opposite so that fermi symmetry alone does not enforce it.  

Let us consider the case of a quadrupole mode of a strongly interacting bosonic gas excited in a spherical trap by imposing an ellipsoidal distortion of it and then releasing it abruptly to ring down to rest.  It is interesting that the modes do seem well characterized by Eulerian hydrodynamics.  This implies that the excited eigenstates of the spherical harmonic trap include such collective motions.  Interestingly, this is not the case for a noninteracting gas.  A single boson will oscillate about the direction of contraction but not in an ``elliptic'' fashion where a pressure field drives expansion in perpendicular directions.  The oscillation frequencies are well described by the Gross-Pitaevskii equation and the pressure field given by the nonlinear interaction among terms.  The G-P equation, the Hartree-Fock equation, Ginzburg-Landau and all other nonlinear effective equations describing electrons are problematic in the sense that we know the dynamics of the underlying many body wavefunction is linear.  These nonlinear equations lump the effects of the other particles into a nonlinear term and this is successful for some purposes but nonlinear equations simply do not give superpositions.  Instead of trying to justify why G-P \textit{must} hold let us see \textit{when} it holds and for \textit{how long}. 

Consider the case of our deformed gas held static and assume it has been somehow cooled to its ground state (ignoring n-body decay processes).  Now we alter the trap abruptly to a spherical shape and let the gas begin to oscillate.  This is expected since it is not in the ground state of this new trap.  Quantum mechanics tells us that it will continue in this superposition for all time.  However, when we look at the trap, specifically we \textit{see} the 1-body density function, it seems to oscillate according to G-P with some damping term that leaves it in a spherically symmetric state.  How can we reconcile these two pictures?  The answer is in that the 1-body density function is a crude measure of the real many body dynamics of the system.  Even though the final state seems spherically symmetric, the Schr\"{o}dinger equation guarantees that there are oscillating currents persisting so that we cannot necessarily compare such a state to a similarly sized ball of classical gas that we cooled down to the same radius in a spherical trap.  The moral of this is that G-P is only expected to be valid for a finite period of time while the n-body correlations are constant.  In other words, if the system is so rigid in its evolution that it only has a 3D continuum of degrees of freedom, then it is a candidate for such effective nonlinear field evolution.  

One might then ask: what is the meaning of the ``order parameter'' $\Phi$ we use to describe it (or equivalently, $\rho$ and $v=\nabla \phi$)?  We can try to peg this to some value of the wavefunction in a coordinate slice or define it by the best 3D wavefunction $\psi$ that gives the net many body $\Psi\approx\prod \psi_{i}$.  This clearly works only for bosons.  
Once the macroscopic ``external'' motion gets channeled into these directions, the 1-body density function will cease to be describable by it.  It is interesting that the interaction potential is effectively repulsive for these experiments and the only attractive force is the external potential that is constraining the flow and moving slower than the internal oscillation speed.  Except for trap release problems this is universally true and may explain why such an order parameter approach can be successful.  Nevertheless, at the trap edges we will see that important breakdowns of this are inevitable and must be considered in questions of damping and angular momentum conservation.  

To illustrate these issues more specifically, consider the case of a rotating elliptical flow.  Such a state is often obtained by exciting a ``scissors'' mode and then releasing it into a spherical trap.  To avoid complications let us consider a state generated by a rotating elliptical trap that has been cooled to the lowest extent possible for this driven case.  Firstly, we don't know that there is any stationary state corresponding to this motion given the trap size and rate of rotation so there may be current oscillations hidden in it.  Now let us release such a cloud into a spherical trap of similar size and let it evolve.  This evolves as an irrotationally evolving cloud that gradually spreads and produces a spherical cloud when viewed in the one-body coordinate space described by $\Phi$.  Based on a classical hydrodynamics, we would argue that the irrotational angular momentum of the cloud has now converted some of its energy to heat and is now in a rigid body rotation.  However, the system is still a many body wavefunction and vorticity can only enter such systems in a singular fashion.  Vortices in many body bosonic gases do not have to show up as depleted density lines in the 1-body density in general.  In the strongly interacting case, this is often true and the reason such states are, rightly or wrongly, labeled superfluid.  Vortices in such systems is typically of a finite lifetime which often begs the question of what happens to the angular momentum associated with them.  In our gas, vortices may never appear at all.  However, the many body wavefunction need not be hyperspherical just because the 1-body density is spherical.  The variations in the many body $\Psi$ allow density waves so that the motion can still be irrotational, conserve angular momentum and not dissipate any of this energy into the kinds of uncorrelated relative coordinate motion we would typically label as ``heat.''  An example of this is discussed in the next section.  

Are there physical manifestations of such final states that distinguish them?  In other words, can the same final 1-body spherical distributions be distinguished in some observable fashion?  Common probes of such clouds involve oscillation rates and interference.  If we deform this new cloud and release then we can investigate the rate of the induced oscillatory modes.  Alternately, we can often delocalize a rotating cloud transversely then bifurcate it along a plane perpendicular to the motion.  By moving these clouds along different paths then recombining them can be interfered and this indicate something of the internal motions.  A rigidly rotating trap will have many internal vortices but similar phase advance along the outside edges.  One with correlated surface waves will have most of its motion at the edges so interference on the interiors should be negligible.  The stiffness of the trap against deformation depends on the distribution of vorticity within it.  Let us consider three cases:
\begin{enumerate}
\item The cloud is still moving with irrotational motion but with the advancing surface deformation no longer evident in the one body density function $\rho(x)$.
\item The cloud has behaved like a classical gas and has settled into a rigid body rotation with the lost kinetic energy going into heating the cloud.
\item The angular momentum has been driven out in the manner of vortices escaping the cloud or into some thin classical gas ``halo'' such that the bulk of the kinetic energy has gone into heat.
\end{enumerate}
We will now consider some examples that suggest when each is a valid picture and motivate experiments to determine when many body properties can persist and thwart otherwise classically inspired approaches.  

\section{Faux Vorticity}
The working assumption of many of those working on ultracold gases is that, after some evaporative cooling and a few rf sweeps, they equilibrate it and then we can expect thermodynamics to hold and, when strongly interacting, hydrodynamics to hold as well.  A favorite workhorse of these problems is the gas in the harmonic trap.  By extending one direction so it is much longer than the others we can create a near axial symmetry and have a 2D (in some senses) system.  By deformation in the transverse directions these can attain an elliptical shape and be driven into rotation.  At critical driving velocity, $\Omega_{c}$, one can attain vortex production typical of superfluid Helium.  A few problems arise with this association and any attempts to apply the two fluid model.  Firstly, these vortices are typically transient and spiral out of the gas cloud in time.  This is a major problem for angular momentum conservation.  It is typically assumed there is a classical ``halo'' about the cloud that absorbs this.  Secondly, the critical driving frequency is much higher than the estimates based on energy and angular momentum from the G-P equation.  The presence of durable vortices at all is suggestive of some irrotationality constraint on the motion and the G-P equation predicts this.  Indeed, it is hard to see how ``rigid body rotation'' can emerge from a many body wavefunction in any classical limit except as a deforming lattice of vortex lines.  The G-P equation naturally predicts complete damping on the density at the center of the core but this is far from true.  Many believe that this is because the core is tilted relative to the viewing angle but, since these details are smaller than the healing length, it is more likely a many body feature not captured by G-P and any macroscopic order parameter.  

Let us now consider a bosonic gas in such a driven elliptical trap that is spun up from rest.  Generally, after some point in the spinning some evaporation occurs and then the spin up continues.  Presumably this is to equilibrate the gas in some fashion but we should wonder how vorticity enters at all.  These are topological obstructions that have to enter as loops or from infinity through the long thin tail of the many body wavefunction.  If these form growing loops then the edges of them presumably loop back near the edge of the cloud where interactions and density become small.  This allows for a path for the visible vortex to get drawn out and dissipate in a topologically consistent fashion.  It also suggests that the angular momentum and energy get deposited at the edge of the cloud.  In this region the interactions are weaker and so the coherence enforced by the G-P approximation fails.  This allows surface waves that are ``uncorrelated'' so can contain energy and angular momentum in a fashion not detectable in the one body density function that we optically project.  

Before we consider this case, let us consider a simpler one without vorticity and that has weak interactions.  The cloud spins in an elliptical fashion then is abruptly released into a cylindrically symmetric trap.  The lack of interactions means the above uncorrelated waves can occur and there is no source of vorticity so the motion is irrotational as a many body wavefunction, contains angular momentum and tends to a cylindrically symmetrical one body density cloud.  This is a complete contradiction from the point of view of classical hydrodynamics.  Since it is the author's thesis that these clouds often contain hidden history dependent information, this is an illustrative example.  There will be a trace of this behavior in the density profile of the cloud so gives a place to investigate this suggestion in current experiments as their resolution improves.  

Consider a cloud with density profile 
\begin{equation}
f(x,y)=1-\left(\frac{x^{2}}{a^{2}}+\frac{y^{2}}{b^{2}}\right)
\end{equation}
rotating at frequency $\Omega$ in an elliptically distorted harmonic trap.  We choose $a>b$ without loss of generality.  
 \begin{figure}
  \begin{centering}
 \includegraphics[width=2in]{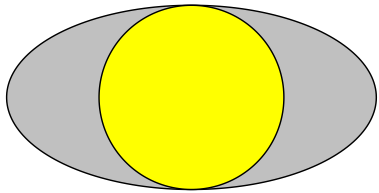}%
 \caption{\label{ellipses}  A top view of the elliptical cloud.  The angular integration is done over $2\pi$ radians in the yellow region and over restricted arcs in the gray support outside it.}
 \end{centering}
 \end{figure}

To get the density profile of the cloud when it has so decorrelated we average the density over each radius.  The integration can be broken up into two distinct regions for the cases $r<b$ and $b<r<a$ as in fig.\ \ref{ellipses}.  The integration is to average over all equal radius arcs over the support of the cloud.  In the first case the integration is over $2\pi$ but the second has it bounded in two arcs bounded by $[-\Theta(r),\Theta(r)]$ where 
\begin{equation}
\Theta(r)=\tan^{-1}\left(\frac{b}{a}\sqrt{\frac{a^{2}-r^{2}}{r^{2}-b^{2}}}\right)
\end{equation}

\begin{align*}
  \rho = \begin{cases}
      1-\frac{1}{2}\left(\frac{r^{2}}{a^{2}}+\frac{r^{2}}{b^{2}}\right) & \text{if $r<b$} \\
      \frac{2}{\pi}\left(\Theta(r)\left(1-\frac{1}{2}\left(\frac{r^{2}}{a^{2}}+\frac{r^{2}}{b^{2}}\right)\right) - 2\tan\Theta(r)\left(1-\frac{r^{2}}{b^{2}}\right)\right) 
      & \text{if $b<r<a$}
    \end{cases}
\end{align*}

For the case of the ellipse with $a=2,\ b=1$, we have the final cloud density, $f(r)$, compared with the densities along the initial correlated gas cloud along the major and minor axes.  The cloud density we have chosen is that of a gas in the Thomas-Fermi limit where interactions are strong enough so that the distribution is described by a truncated parabola \cite{Stringari}.  This provides an upper limit on the possible behavior not a rigid illustration for all interaction strengths.  It probably has best validity for dilute gases with large enough clouds so the edge resolution is not that important and elliptical distortions that are large.  It is interesting that there is a nonanalytic feature at $r=b$.  There is no reason to expect such a sharp delineation in a real cloud but the profile structure is also distinctive in its nonmonotonicity and is a likely indicator of such decorrelated edge waves that are hiding both angular motion and radial motion that are not describable by a hydrodynamic picture.  It is interesting that such angular motion can be irrotational in that it arises from a vortex free many body wavefunction yet the time and averaged motion seems to be rotational.  In this sense we refer to it as ``faux vorticity.''  For this reason we should generally be cautious of using such averaged velocity fields for dynamics.  
 \begin{figure}
  \begin{centering}
 \includegraphics[width=2in]{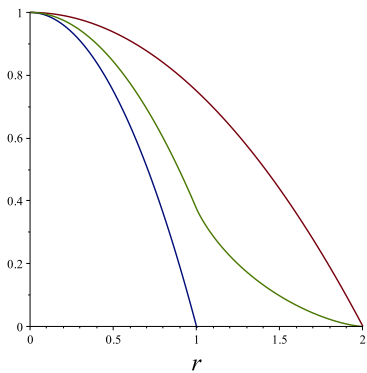}%
 \caption{\label{profile}  Final cloud density profile as a function of radius (green) vs.\ the major and minor axis densities for the initial cloud (blue and red). }
 \end{centering}
 \end{figure}

We can explicitly calculate the velocity in this case.  
The velocity field is irrotational and begins as $v=\nabla\Phi$ where \cite{Lamb45} 
\begin{align}
\Phi=\Omega x y \frac{a^{2}-b^{2}}{a^{2}+b^{2}}.
\end{align}
The angular momentum is nonzero and related to the angular momentum of the solid body by
\begin{align}
L=L_{sb}\frac{a^{2}-b^{2}}{a^{2}+b^{2}}.
\end{align}
Once the cloud delocalizes the mean velocity field can be found by averaging over all angles.   
The radial component vanishes and the nonzero component arises from the the gray region in fig.\ \ref{ellipses} so that 
\begin{align}
v_{\theta}=2\int_{-\Theta}^{\Theta}\Omega r^{3}\frac{a^{2}-b^{2}}{a^{2}+b^{2}}\cos(2\theta)d\theta.
\end{align}
Hence 
\begin{align*}
  v_{\theta} = \begin{cases}
      0 & \text{if $r<b$} \\
          \frac{\Omega b^{2}}{\pi}\left(\tan\Theta(r)\frac{1-\frac{r^{2}}{b^{2}}}{\frac{a^{2}}{b^{2}}-1}
          \right) 
      & \text{if $b<r<a$}
    \end{cases}
\end{align*}
which is certainly not irrotational.  For our sample ellipse with $a=2, b=1$ we have an integrated vorticity of $\int \nabla \times v=\frac{\Omega}{6}$.  This is just one way in which using spatial averages we can arrive at unphysical conclusions for a semi-classical approach.  In fact, it can be shown that the use of hydrodynamic variables for such a general many body quantum gas leads to profoundly nonlocal equations of motion \cite{Chafin-II}.  

For such a cloud one now can ask what sorts of changes in the frequency modes and damping rates occur.  If ``thermalization'' was really a property of such gases there would be no such question.  The number of particles and their net internal energy would determine the cloud shape and all such properties.  The mode frequency is a measure of the mass moving by displacement and the restoring force.  Since this new distribution has a greater distribution at the center and a thin tail at the extremes we expect the mode frequency to be reduced.  This is because there is a larger fraction of mass moving in a weaker potential.  Damping rates are measured through the one body density equilibration time.  However, if we view the cloud as a set of phase shifted rotating clouds then when we distort the axial symmetry of the trap quasi-statically towards its original shape and then abruptly release, the distribution of phases has been altered so that the system is back to its original state and the evolution of the modes and damping times are the same.  It seems that the stored memory in such an interacting cloud is important in its dynamics and that hydrodynamics may be a poor starting point for general initial data.  It should be noted that a gas driven to rotation by a rotating elliptical trap has far off nonrotating tails that necessitate the introduction of vorticity here i.e. the ideal irrotational motion of the cloud \cite{Stringari} is not realistic.  This is important because such an assumption is of profound difficulty for viscosity based models of damping of such clouds \cite{Schaefer} and gives no where for the angular momentum to shed in damping other than through a vague classical ``halo'' of gas.  

By previous arguments such a gas must be a delocalized intrinsically quantum object.  The generally small size and thin optical depth of such gases makes their equilibration through photon production impossible on any reasonable time scale.  The evaporative cooling methods work to limit the spread of the energy eigenstates in the cloud, $\Delta E$.  Since this is done while spinning the trap this suggests there is little relative internal current to generate photon production as well.  When the cloud is released into an axially symmetric trap on is confronted with the additional problem that rotating eigenstates do not have angular density variations.  This is generally true in the many body case as it is for Hydrogen atom ones.  The consequence is that such a released cloud has as $\Delta L$ spread that is large enough to form the distorted cloud we initially observe.  Phase effects invariably lead to a loss of coherence in the one body density function.  It is this author's opinion that this is the true source of ``damping'' and that it is of a fundamentally different nature than classical gases.  The local velocity fields we associate to the cloud by averaging the phase shifted irrotational motions is rotational and thus N-S hydrodynamics predicts a local damping by thermal momentum transfer across the shear lines.  Vorticity is pulled in or pushed out through the gas surface.  In our trapped gas, however, there is little thermal momentum flux and the true many body velocity flow is irrotational (in that it comes from the gradient of the many body phase $\tilde\nabla \Phi(X)$).  The system is ``vorticity poor'' and angular momentum gets trapped in decorrelated surface waves at the edge of the cloud which should provide another hidden source of history for the cloud dynamics.  It may even pave a way to explain the tendency of vortices in strongly interacting gases to drift out of the cloud with apparent loss of angular momentum and energy.  

\section{Vortex Buoyancy}
Let us now consider the case of angular motion of a strongly interacting gas which forms a vortex lattice.  This seems so persistent that it is tempting to consider it to be an angular momentum eigenstate of a gas.  We can see that this cannot be true, no matter how much we cool it, because the currents are not constant in time.  When we observe the current at a given point in the center of mass frame the vortices advance in a rigid body fashion and the current vary as the vortices pass.  We realize that the vortex lattice can also be shifted by any angle as valid initial data.  Certainly superpositions of such states would give blurred or indistinguishable lattice locations.  This situation is reminiscent of the observations of special states of macroscopic classical matter \cite{Chafin-I} so they must be superpositions of eigenstates.  The visible density varying structure is an indication that we have imprinted some aspect of our macroscopic classical world on the system and that it can be expected to decay with time.  The rate of change of these currents gives an indication of the spread in energies, $\Delta E$, in the eigenstates used in its construction.  It seems unavoidable that there is also a spread in the angular energy, $\Delta L$, of the eigenstates as well to produce the angular localization in the observed gas shape. 

Now let us consider the following thought experiment.  Let us take this lattice and turn down the interaction so the vortices disappear then return it to its original value.  Do the vortices reappear?  I argue that they do not because the phase information has been lost, which was only transient anyway.  The angular momentum and overall vorticity (since it is a topological invariant) is still present in the cloud so we either have a superposition of correlated vortex lattices or we have some other more complicated and perhaps less rigidly correlated.  It is the uniform distribution of vorticity that characterizes both uniform lateral shear and rigid body motion.  The first case is dissipative and generally requires surface vorticity contributions at the boundary of the fluid.  Rigid body rotation seems like the only long term alternative but we have already seen that quantum effects can result in correlated motions that thwart hydrodynamics.  In the case of cold gases the vorticity is accompanied by an exclusion of density around the vortex that is typically determined by the healing length, thermal wavelength, fermi wavelength or some other length scale characterizing the particular gas.  For bosons at temperatures where $\lambda_{deB}$ is much smaller than the cloud size and the interparticle separation this may lead to important effects.\footnote{We have implied that the oscillations here are clustered about a typical size so that such a $\lambda_{deB}$ is descriptive.  When the excitation energy per particle is higher than the interaction energy between the particles in the ground state, this is possible but not when $\Delta E$ is too large.}

Firstly, the role of vorticity in creating angular momentum is often oversimplified.  A vortex with single winding number of phase is typically said to have angular momentum $L=\hbar$.  This is certainly true for all the H-atom wavefunctions and any symmetric cloud where the vortex is centrally located or when we are measuring the vorticity about the center of the vortex (which naturally presumes the vortex core is a line).  A vortex that has moved to the edge of the cloud generally generates less than $\hbar$ of angular momentum.  In the case of a strongly interacting cloud, where the electrons all must cooperate, the momentum is $N\hbar$ in the symmetric case.  For clouds with intermediate angular momenta it is a mystery how this gets arranged in a stable fashion.  Certainly there are states of such intermediate angular momentum.  The Gross-Pitaevskii equation has an analogous problem in producing a gapped spectrum of excited states (which are forbidden by the Hugenholtz-Pines theorem \cite{HP}).  The resolution to such problems is that there has to be a distinction between standing waves in hydrodynamics and constrained dynamics like G-P and the eigenstates of the underlying many body wavefunctions.  It must be remembered that eigenstates in fixed traps have either no current or uniform currents so there is not a 1-1 correspondence between these and the hydrodynamic modes.  The vortex lattices we observe in such traps and liquid Helium are not angular momentum eigenstates of the many body wavefunction for this reason, even if they are cooled to absolute zero.  In principle one could wait long enough for these lattices to angularly delocalize and appear as a uniform blob with no angular density variation if the system was strongly isolated.  This angular momentum is, of course, still present but not describable by a G-P state.  Such intermediate angular momentum states are described by vortices not centrally located in the cloud or by such hidden correlated motions as in the previous section.  It begs the question of what would happen if one cycled the interaction strength of a vortex containing gas in an axially symmetric trap.  If one starts with a vortex lattice, does it return?  If not, is the angular momentum lost?  

In the case of a strongly interacting gas one has vortices on the scale of the healing length $\xi=(4\pi n a)^{-1/2}$ where $n$ is the particle number density and $a$ is the scattering length.  In the case of a weakly interacting gas, vortices still exist with scale given by the typical or thermal wavelength $\lambda_{deB}$.  In both cases there is a volume exclusion driven by the vorticity itself.  When angular momentum is given by surface waves or other correlated edge motions, this is not required but even this requires some depletion in density as was shown in the previous section.  For these reasons we must conclude that angular momentum itself carries with it a buoyancy.  This is unlike heat, which, while it produces an expansion, as $\Delta E$ of the state becomes negligible, does not tend to produce higher temperatures at higher altitudes in a potential.  At room temperatures where $\lambda_{deB}$ is not just smaller than the interparticle separation, but much smaller than the size of the atoms themselves,  one would need many vortices between neighboring atoms to have an appreciable effect.  Interestingly, this is near where the shear becomes significant on the scale of the particle separation.  Such a high Knudsen number regime has always been problematic for perturbative expansions of the Boltzmann equation \cite{Cohen}.  Some argue that this is because fluctuations become dominant here but such a discussion does not work for a quantum picture of a gas with very small $\Delta E$ so that a more quantum picture of high shear gas flows may provide a better resolution.

Our purpose here is to investigate the role of this buoyancy on clouds in traps.  We are not simply concerned with highly interacting regimes, where the coherency of the one body density persists for long times, but the case of weakly interacting gases and ones with high internal energy and spreads of energy states that can be considered persistently athermal.  Strongly interacting gases can produce rigid lattices with Tkachenko oscillations.  We are interested in the limit where vorticity is completely uncorrelated in that no such organized pattern of vortices exists or condenses into a pattern in the one body density function and the forces between such vortices are small.  Since the case of a harmonic trap introduces some significant mathematical complications, let us start a case of a rotating cylinder of gas that is both weakly interacting and dominated by a single wavelenth $\lambda$.  The interaction with solid walls makes us worry about localization effects associated with quantum measurement there \cite{Chafin-pip-meas}.  For this reason let us consider an optical trap with sharp enough edge potentials to create this cylinder.  

The hydrodynamic limit follows directly from first order perturbations of the Boltzmann equations.  Higher perturbations arise as the Knudsen number grows.  For the case of a rotating cloud, it is not surprising that as the edge velocity grows as a fraction of the thermal velocity or speed of sound that we also see corrections.  We need these first before we can see the effects of vortex buoyancy.  The static equilibration condition balances the pressure $P=nkT$ forces with the centrifugal ones so that 
\begin{align}
\rho\frac{v^{2}}{r}=\frac{kT}{m}\frac{d\rho}{dr}
\end{align}
thus
\begin{align}
\rho(r)=\rho_{0}e^{\int_{0}^{R}\frac{m}{kT}\frac{v^{2}(r')}{r'}dr'}.
\end{align}
Parameterizing the velocity field as $v=U(r+Vr^{2})$ and fixing the net mass, angular momentum and total energy of the gas
\begin{align}
M&=2\pi\int\rho(r)rdr\\
L&=2\pi\int\rho(r)v(r)r^{2}dr\\
E&=2\pi\int\left( \frac{1}{2}\rho(r)v(r)^{2}+\frac{3}{2}\frac{kT}{m}\rho(r) \right)
\end{align}
we can reduce the parameters $U,V,T,\rho_{0}$ down to one degree of freedom.  To complete the solution we need to maximize the entropy.  This can be computed as an integral of a local quantity, but we expect some global condition to drive the system.  Entropy seeks to maximize the internal energy of the gas.  This is the same as minimizing the macroscopic energy (kinetic energy in this case).  We are interested in measuring the deviation from rigid body motion so define the rotational frequency $\Omega$ implicitly by $L=\frac{1}{2}MR^{2}\Omega$.  In terms of this, the resulting velocity and density (mass per area) profile are
\begin{align}
v(r)&=\Omega r\left(\left( 1+\frac{2}{7\cdot3} \frac{(\Omega R)^{2}}{v_{th}^{2}} \right)-\frac{5}{7\cdot6} \frac{(\Omega R)(\Omega r)}{v_{th}^{2}}\right)\\
\rho(r)&=\frac{M}{\pi R^{2}}\left( 1- \frac{1}{12}\frac{(\Omega R)^{2}}{v_{th}^{2}}+\frac{1}{6}\frac{(\Omega r)^{2}}{v_{th}^{2}}\right)
\end{align}
to low order where $v_{th}=\frac{3kT}{m}$ is the rms velocity of the gas.  The discrepancy with the rotating case is in the central density $\rho_{0}$ but most dramatically in the equilibrium velocity field which now has a backwards $\Omega r^{2}$ component.  These effects are small relative to the speed of the gas particles/sound.  In solids, the speed of sound is an upper bound on the cohesive energy of the body.  In a harmonically trapped gas there are critical velocities for vortex production (when strongly interacting) and for when the rotational forces simply overwhelm the trap.  In our case, these are not a concern but we still have the dramatic result that rigid body motion is not the equilibrium limit.  This means that no perturbative modification of hydrodynamic viscosity can be expected to be accurate since all of these lead to rigid body motion as their final result.  

Now let us consider the effect of this density depletion at the center on the flow.  For higher T gases we expect incoherent vortices (where $a\ll\lambda_{deB}<d$ where the atomic size is $a$ and particle separation is $d$) unlike the kind of observable ones we see in images of vortex lattices.  There is no reason to suspect that this leads to any pattern in the one body density of the wavefunction of the gas at all, however, the buoyancy of such vortices means they should accumulate in the center and accentuate the parabolic density profile of the trap.  This gives an equilibration condition
\begin{align}
\rho'\frac{v^{2}}{r}=\frac{kT}{m}\frac{d\rho}{dr}
\end{align}
where $\rho'$ is the density reduced by the vortex density of the flow and $\rho$ is the ``background'' density where the vortices are absent.  Using the circulation condition for quantum vorticity and that the velocity field $v=v(r)\hat\theta$ we have that the number density of vortices is
\begin{align}
n_{v}=\frac{m}{2\pi\hbar r}\left( v(r)+r\frac{dv}{dr}  \right).
\end{align}
The area excluded by each vortex is $\beta\lambda^{2}$ where $\beta\sim\mathcal{O}(1)$.  This leads to a reduced density $\rho'=\rho(1-(\beta\lambda^{2})n_{v})$.  The resulting velocity field is unchanged to this order but the density profile becomes
\begin{align}
\rho'(r)&=\frac{M}{\pi R^{2}}\left( 1- \frac{1}{12}\frac{(\Omega R)^{2}}{v_{th}^{2}}(1+\delta)+\frac{1}{6}\frac{(\Omega r)^{2}}{v_{th}^{2}}(1+\delta)\right)
\end{align}
where $\delta=\frac{\beta\Omega\lambda^{2}m}{\pi\hbar}$.  This is typically a small effect in the given configuration until the frequency of the thermal oscillations becomes comparable to that of cloud rotation.  For the case of harmonic traps, where the edges are much less dense, it should be reversed and greater in magnitude.

\section{Conclusions}
Our discussion on the origin of thermalization in gases depends on photon production and the ability for temporal fluctuations to drive prolific low energy photon production.  For initial data with large enough fluctuations there is reason to believe that these fluctuations drive amplitude uniformly through the many photon number spaces and creates the kinds of distributions derived by the microcanonical ensemble.  This provides a path to ``autothermalization'' that is not dependent on coupling to an external universe.  Given the very formal character of quantum statistical mechanics which leaves one grasping for how to treat nonequilibrium dynamics and the long list of open problems here (e.g. nucleation theory which is one of the most unsuccessful areas of theoretical physics), this is a significant step forwards.  It is interesting that this author's approach to quantum measurement also depended on cheap and prolific photon production and that this is generally ignored in decoherence approaches and the eigenstate thermalization models.  The paradoxes of measurement and equilibration are typically always derived in spaces where the photon number is implicitly fixed.  

In the case of the hydrodynamics of ultracold gases, we are somewhat hampered in our approaches in that we have never arrived at a completely quantum theory of gases in the classical domain.  Current approaches are a kind of hodge-podge of quantum thermodynamics and classical hydro.  The perspective here is that the system is a many body wavefunction at all times rather than a mixed state.  In the ultracold gas, we have assumed that the gas is so optically thin and the coupling to radiation so small that there is no important thermalization or spreading of photon numbers in the durations of the experiments.  Such a situation means that there is no avenue to reduce $\Delta E$ of the eigenstate distribution so athermal states can persist.  

We have investigated rotating gases in two extreme cases: highly quantum-like where vorticity is expensive and very classical where it is cheap.  The first gives decorrelated many body edge motion hidden from the one body density function and leads to a fictitious vorticity in its velocity.  The second displays quantum corrections due to high vorticity density in a regime that would generally be considered to be classical.  Even in this classical limit we saw there are corrections to the rigid body rotation curve and the quantum effects enhanced it.  For the case of harmonic traps with dilute edges, such an effect should be more pronounced.  In the limit of high vortex density i.e. the regime where the Barnett approximation is problematic for classical gas theory, this suggests a quantum gas approach may be essential to treat it properly.  This is not a a resolution of the convergence problem for classical kinetics but it is interesting that it is a domain where classical kinetics may not apply.  It would be very interesting if such a quantum treatment led to a perturbative treatment of gas dynamics without convergence problems.  It has been assumed that temperature is a global constant in these traps.  Given that we found some variation in the rigid body rotation curve, it may be prudent to do more sophisticated calculations to verify this is in fact so.  


\end{document}